\documentclass[%
 reprint,
 amsmath,amssymb,
 aps,
]{revtex4-2}

\usepackage{blindtext}
\usepackage{hyperref}

\usepackage{graphicx}% Include figure files
\usepackage{dcolumn}% Align table columns on decimal point
\usepackage{bm}% bold math
\begin{document}

\preprint{APS/123-QED}

\title{The collective statistical mechanical personality of a group}

\author{Sarah Marzen}
\affiliation{Department of Natural Sciences, Pitzer and Scripps College}
 \affiliation{Kravis Department of Integrated Sciences, Claremont McKenna College}
 \affiliation{National Institute for Theory and Mathematics in Biology}
 \email{smarzen@natsci.claremont.edu}

\date{\today}% It is always \today, today,

\begin{abstract}
We propose a mathematical framework for organizational psychology based on a Maximum Entropy model of a group's personalities. The Maximum Entropy model is then decimated to a single ``collective personality''. If the original personality scores are augmented by intelligence and emotional quotients, then a collective intelligence is also mathematically revealed. With simple matrix analyses of the collective personality, one can understand: that weak interpersonal coupling can strongly affect group character; that malleable rather than stubborn personalities control the group's collective personality; that one can mathematically solve for optimal top-down directives to achieve certain group personalities; and that groups can have a personality disorder even if the individuals composing the groups are neurotypical. We hope that this framework provides a useful starting point for future mathematical analyses in organizational psychology related to innate character rather than opinion dynamics or decision making, and note that the analysis can be applied to much more complex Maximum Entropy models than the one proposed here if empirical evidence suggests that the Gaussian model proposed here is overly simplistic.
\end{abstract}

\maketitle

%\tableofcontents

Groups and organizations have become more and more important in society, injecting information and decisions collectively rather than as individuals with similar top-down directives. The study of organizational psychology tracks the performance of groups and looks for factors that determine the performance of groups \cite{barrick1991big,barrick1998relating,woolley2010evidence}. We propose here the radical idea of ``decimating'' the group to a single person whose characteristics are determined by intrapersonal and interpersonal correlations in personality and even intelligence.

Such underpinnings give rigor to ideas that are often expressed colloquially. When Mitt Romney ran for President, he famously said, ``Corporations are people, too.'' The New York Times responded, ``If corporations are people, then corporations are psychopaths.'' We can take this farther. One might contend that most charities seem to be communal narcissists, constantly touting their prosocial successes. In fact, perhaps very few organizations appear to be psychologically normal when viewed as a collective.
But what do we mean when we say these things?

In this paper, I attempt to make such statements precise by introducing a simple Maximum Entropy model of groups of people in which individual people are represented by five-dimensional vectors that encode one's Big Five personality, magnetic fields correspond to top-down directives from leadership, and coupling constants correspond to how two people interact with one another. The dimensions of each person's vector are inexorably intertwined, as my introversion may interact negatively or positively with your neuroticism. Using the typical decimation procedure for such statistical mechanical models of averaging, one can find a single vector that describes the collective personality of the entire group of people. Depending on the resultant coupling constant and magnetic field after decimation, the collective personality maps onto different and well-understood personality disorders \cite{samuel2008meta}, with the rare organization managing to be both stable and psychologically normal.

The minimal model of a group of people and the decimation procedure are subject to fine-tuning, of course. Simplicity in the Maximum Entropy model is not required for the idea of decimation revealing the collective personality to be valid. The key idea is that once a Maximum Entropy model has been made of the personalities in an organization, using ideas about how people can be represented in terms of personality by their Big Five scores, decimation yields the collective personality of the group, as well as its variability around its mean Big Five personality score.

These models, if accurate, might have some utility in enabling the C-suite to improve top-down directives: Different top-down directives result in different magnetic fields, which then result in different collective personalities. If top-down directives are optimized, a stable and optimal collective personality (a constant desire of corporate retreats) results. And extensions of the Maximum Entropy model and subsequent analysis proposed here allow for a mathematical grounding of the idea of the ``c-factor'', or the collective intelligence \cite{woolley2010evidence}.

What follows is a short example of how such a procedure can be used to understand groups and organizations whose Maximum Entropy model is Gaussian. The basic ideas behind understanding collective personalities from Maximum Entropy models, however, apply far beyond the Gaussian case-- though analysis of the Gaussian case does lead to interesting insights.

\section{Background}

This manuscript starts by proposing a minimal model of how personalities in a group interact with each other, with natural extensions to how personalities combined with intelligences interact with one another.
We are taking something incredibly complicated and making it overly simple, with the ability to complexify at will by choosing a more complex Maximum Entropy model. Maximum Entropy models are minimally biased models in that they have maximal entropy given various constraints. If constraints are chosen well, very simple Maximum Entropy models can describe very complex processes \cite{meshulam2025statistical,schneidman2006weak,bialek2012statistical,bialek2014social,lee2015statistical}, although sometimes one can play the game of adding constraints to Maximum Entropy models to improve the fit \cite{tkavcik2013simplest}. What model should be chosen will depend on what models data well, following efforts to use Maximum Entropy models in neuroscience \cite{meshulam2025statistical}, bird flocking \cite{bialek2012statistical,bialek2014social}, and even Supreme Court decisions \cite{lee2015statistical}.

That being said, many researchers have tried to model related phenomena with simple quantitative models that capture the essential behavior. These models are all related rather than identical in what they are modeling. Opinion dynamics models \cite{shirzadi2025opinion,castellano2009statistical} model the opinions that people can hold-- their thoughts rather than their overall personalities. Meanwhile, group decision-making models \cite{kugler2012groups} model the decisions that a group makes, the process of which is absolutely related to group personality but is not determined by group personality. There are a range of opinion dynamics models and group decision-making models, such as the voter model \cite{castellano2009statistical} or Bayesian models \cite{dong2024cognitive}. Some of these models take a similar mathematical form to the Maximum Entropy model used here. For instance, the Curie-Weiss model has been used to model opinion dynamics as an approximation to a model in which opinions are coupled ``spins''. But the mathematical similarity does not imply that the phenomena modeled are identical or that the same analyses are useful.

Many of the models employed to study opinion dynamics or group decision-making are dynamic in nature and describe a sequence of opinions or a sequence of decisions \cite{castellano2009statistical}. In contrast, the model of personalities described here is static, with the idea that while opinions and decisions can fluctuate wildly from second to second, personalities once in a group are stationary. We simply mean that there is an innate component to personalities that drives intrapersonal and interpersonal coupling, as psychologists roughly believe that personality stops changing in adolescence with some caveats \cite{slobodskaya2021personality}. The stationary condition means that the distribution of personality scores that we might receive does not vary as time changes.

The Maximum Entropy model proposed is a minimal mathematical model that describes something that is within the realm of organizational psychology. The models used there are not like the minimal model described here and instead are regression-based. Still, the organizational psychology literature provides guidance as to what a minimal model might look like. The Big Five personality traits were shown to be relevant in organizational contexts \cite{barrick1991big}, and the quantities proposed in Sec. \ref{sec:results} as capturing collective personality were inspired by prior mean and variance analyses \cite{bell2007deep,barrick1998relating}. Similar ideas, that collective properties of groups emerged from correlations between individuals, have guided non-mathematical literature on collective intelligence \cite{woolley2010evidence}. But no mathematical model that underpins collective personality or intelligence exists in organizational psychology, which our Maximum Entropy model (or any Maximum Entropy model based on Big Five personality scores and potentially intelligence and emotional quotients) provides.

\section{Setup}

Commonly, to assess personality, subjects take a personality test and get a ``Big Five'' personality score. The test is composed of statements that are assessed on a Likert scale, where the truth of the statement is rated between $1$ and $5$, with only integer ratings allowed. Several questions pertain to each of five personality dimensions: openness to new experiences, meaning a preference for novelty over routine; conscientiousness, related to self-discipline and reliability; extraversion, meaning that one has positive energy in social settings; agreeableness, meaning that one has empathy and avoids conflict; and neuroticism, related to emotional instability and anxiety. The several questions that assess each personality dimension combine to make one score in each personality dimension. These scores are often percentiles or related to percentiles, such that higher scores mean that you are more likely to show evidence of that trait. So, a high neuroticism score means that you measure as more neurotic.

For our purposes, we find these personality tests useful because the Big Five personality score reduces an entire personality to a five-dimensional vector with real numbers that are between $0$ and $100$. Personality is, of course, more complicated than a Big Five personality score, but this is a gold standard in psychology for assessing personality, including identifying people with personality disorders.

Taking our cue from the Big Five personality tests, we summarize each person in a group with a five-dimensional real-valued vector $\vec{s}_i\in\Re^5$. (An inverse sigmoid can be applied to normalized percentiles to reveal a real number between negative and positive infinity.) We imagine, in future experiments, that each person in a group takes several personality tests, each one revealing a slightly different Big Five personality score, each score mapping to a slightly different five-dimensional real-valued vector. Without loss of generality, the ordering of traits in the vector will be: openness, conscientiousness, extraversion, agreeableness, and neuroticism.

A simple but effective personality model of the people in a group is likely to be the Gaussian model. In this Maximum Entropy model, we measure and constrain the mean personality scores $\langle \vec{s}_i\rangle$ of every person, but also measure and constrain intrapersonal and interpersonal \cite{woolley2010evidence} fluctuations in scores using variances $\langle \vec{s}_i\vec{s}_i^{\top}\rangle$ and covariances $\langle \vec{s}_i\vec{s}_j^{\top}\rangle$ for $i\neq j$, respectively. This results in a probability distribution over personalities $\vec{s}_i$ given by
\begin{equation}
p(\{\vec{s}_i\}_{i=1}^N) = \exp(E(\{\vec{s}_i\}_{i=1}^N))/Z
\end{equation}
where the energy function is
\begin{equation}
E = \sum_{i,j} \vec{s}_i^{\top} J_{ij} \vec{s}_j + \sum_i \vec{h}_i^{\top}\vec{s}_i
\end{equation}
and $Z$ is the normalization factor that ensures that $1 = \int p(\{\vec{s}_i\}_{i=1}^N) \prod d\vec{s}_i$. The coupling constants $J_{ij}$ can be strong, indicating that this trait in this person influences this other trait in this other person, or small, indicating that there is very little interaction between two traits. Magnetic fields $\vec{h}_i$ are partly based on an innate drive towards a particular personality $\vec{h}_{i,int}$ and partly based on top-down group directives $\vec{h}_{i,ext}$:
\begin{equation}
\vec{h}_i = \vec{h}_{i,int} + \vec{h}_{i,ext}.
\end{equation}
This overall Gaussian model can be thus summarized either by a block matrix of connectivities and a concatenated magnetic field. However, it is useful to see how an organization structures interactions $J_{ij},~i\neq j$ and directives $\vec{h}_{i,ext}$, and so a full understanding of organizational psychology based on this model is able to tease apart $\vec{h}_{i,ext}$ from just what is measured $\vec{h}_i$. If people's personalities are measured in isolation, away from the structure of the group, this teasing apart of innate personality from top-down directives is possible.

\section{Results}
\label{sec:results}

This Gaussian model can be solved explicitly. If we concatenate every person's personality into a really large vector $\vec{S} = \begin{pmatrix} \vec{s}_1 \\ \vdots \\ \vec{s}_N\end{pmatrix}$, the Maximum Entropy model reads
\begin{equation}
p(\vec{S}) \propto \exp(\vec{S}^{\top}\textbf{J}\vec{S}+\textbf{H}^{\top}\vec{S})
\end{equation}
where $\textbf{J}$ is the negative definite block matrix whose individual blocks are all of the $J_{ij}$s and $\textbf{H} = \begin{pmatrix} \vec{h}_1 \\ \vdots \\ \vec{h}_N \end{pmatrix}$ is the concatenation of all the individuals $\vec{h}_i$ in order. Then, from completing the square, we have that the mean $\vec{S}$ is
\begin{equation}
\langle\vec{S}\rangle = -\frac{1}{2}\textbf{J}^{-1}\textbf{H}
\end{equation}
and the fluctuations in personalities are
\begin{equation}
\langle\vec{S}\vec{S}^{\top}\rangle - \langle\vec{S}\rangle\langle\vec{S}^{\top}\rangle = -\frac{1}{2}\textbf{J}^{-1}.
\end{equation}
Note that $\textbf{J}$ must be a negative definite matrix.

Decimation yields the collective personality. For the Gaussian model, this merely means that we average all of the individual personalities. For this simple model, the collective personality has a closed-form solution. More precisely, with the collective personality being
\begin{equation}
\vec{x} = \frac{1}{N}\sum_i \vec{s}_i,
\end{equation}
we have a mean collective personality of
\begin{equation}
\mu = \langle \vec{x}\rangle
\end{equation}
and fluctuations about that mean collective personality given by
\begin{equation}
\Sigma = \langle \vec{x}\vec{x}^{\top}\rangle - \mu\mu^{\top}.
\end{equation}
We construct the averaging matrix $A$, designed so that
\begin{equation}
\vec{x} = P\vec{S},
\end{equation}
which means that $P=\frac{1}{N}\mathbf{1}^\top\otimes\mathbf{I}_5$ is a $5\times 5N$ matrix that repeats the $5 \times 5$ identity block and normalizes by $1/N$. With this averaging matrix in tow, we find that
\begin{equation}
\mu = -\frac{1}{2}P\textbf{J}^{-1}\textbf{H},~\Sigma = -\frac{1}{2}P\textbf{J}^{-1}P^{\top}.
\label{eq:meanS}
\end{equation}
From the original Maximum Entropy model emerges a closed-form expression for the mean collective personality and fluctuations thereof.

While it's nice to have closed-form expressions, these admit no obvious interpretation as is. Therefore, we proceed by making a few reasonable assumptions that could be used when constructing the original Maximum Entropy model about how people interact. We assume, for the rest of Sec. \ref{sec:results}, that interpersonal interactions are homogeneous-- that somehow, the organizational structure of the group supports the same $J_{ij},~i\neq j$ for every pair of people. To simplify notation, we define
\begin{equation}
A_i := J_{ii}
\end{equation}
and
\begin{equation}
B := J_{ij},~i\neq j.
\end{equation}
Then, using the Woodbury identity to simplify the inverse of $\textbf{J}$, we find that
\begin{equation}
\mu = -\frac{1}{2} (\mathbf{I}_5+N\langle\tilde{A}^{-1}\rangle B)^{-1}\langle\tilde{A}^{-1}\vec{h}\rangle
\end{equation}
and
\begin{equation}
\Sigma = -\frac{1}{2N^2} (B+\frac{1}{N}\langle (A-B)^{-1}\rangle^{-1})^{-1}.
\end{equation}
Averages are taken over people in the group, so that $\langle (A-B)^{-1}\rangle = \frac{1}{N} \sum_i (A_i-B)^{-1}$.
In order for the Woodbury identity to be useful, several conditions have to be met, including that $A_i-B$ and $B$ are all invertible. It is very likely that interesting results are possible in the likely case that $B$ is not invertible. But for the remainder of Sec. \ref{sec:results}, we will assume that the Woodbury identity can be applied. See Appendix \ref{app:Woodbury} for details of the calculation.

Another way of viewing these formulae is through the lens of renormalization group flow. This would be a more useful framing for non-Gaussian models, e.g. a variant of a spin-glass Ising model. Essentially, we are decimating to the point that only one ``spin'' is left, with its effective coupling constants and magnetic field. In the limit of a large number of people, we are finding the fixed points of renormalization group flow; for finite groups, we are midway through a renormalization group flow towards a fixed point. Thus, the collective personality is dictated by what ``phase'' we are in. While this interpretation is interesting, it is difficult to extract understanding from known renormalization group flow solutions that typically assume what would be overly simplistic interpersonal and intrapersonal coupling structures and innate personalities. This assessment does not preclude the renormalization group flow viewpoint from being interesting to explore for specialized groups whose minimal models are simple enough that their collective personalities emerge from well-understood renormalization group flows. But in the interests of being practical, we now focus on conclusions of the Gaussian model that would be measurable in organizational psychology experiments and would be of potential interest to organizational psychologists, meaning that we focus on the completely general case of any intrapersonal and interpersonal couplings.

\subsection{Weak interpersonal coupling matters}

In the context of neurons, it was famously shown that weak coupling can still lead to large collective effects \cite{schneidman2006weak,van2025coding}. The same is true with weak coupling in the context of group interactions. If we assume that the interpersonal coupling constants decay as the group grows larger as
\begin{equation}
    B = \hat{B}/N,
\end{equation}
approximating the idea that in a larger group, you have proportionally fewer interactions with each group member, we find that in the large $N$ limit:
\begin{equation}
\mu \rightarrow -\frac{1}{2}(\mathbf{I}_5+\langle A^{-1}\rangle\hat{B})^{-1}\langle A^{-1}\vec{h}\rangle
\label{eq:mu2}
\end{equation}
and
\begin{equation}
\Sigma\rightarrow -\frac{1}{2N}(\hat{B}+\langle A^{-1}\rangle^{-1})^{-1}.
\label{eq:Sigma2}
\end{equation}
In the limit of no coupling, $\hat{B}\rightarrow 0$, we find that the mean personality and fluctuations around the mean personality become:
\begin{equation}
\mu_0 = -\frac{1}{2}\langle A^{-1}\vec{h}\rangle,~\Sigma_0 = -\frac{1}{2N}\langle A^{-1}\rangle.
\end{equation}
There is an appreciable difference between $\mu$ and $\mu_0$, and appreciable difference between $\Sigma$ and $\Sigma_0$, meaning that weak coupling in an organization still leads to measurable changes in mean collective personality and in fluctuations around that mean collective personality. The decaying interpersonal coupling constants $B$ are still relevant to group behavior as long as $\frac{B}{N}$ is $O(1)$.

Note that if interpersonal coupling is not $O(1/N)$ but is instead $O(1)$, meaning that interpersonal interactions retain their strength as the group grows in size, then the collective personality vanishes towards the median possible personality. In other words, strong coupling causes a washout of both innate and external drives. In addition, the fluctuations in personality given by $\Sigma$ are dominated by $B^{-1}$ rather than $A^{-1}$, meaning that the interpersonal coupling rather than intrapersonal coupling dictates fluctuations in personality test results collectively in this limit. But these fluctuations do not give the overly connected group a personality; the fluctuations still decrease as $1/N$. This is potentially a desirable outcome if alternatives are collective personality disorders, as discusssed in Sec. \ref{sec:personalitydisorders}, but this limit requires longer workdays as the organization grows to retain the same number of interactions between disparate individuals as the organization grows in size.

\subsection{Malleable people, not stubborn people, sway the collective personality}

Who shapes an organization? In opinion dynamics models, there is the idea that ``stubborn voters'' can swing an entire election \cite{yildiz2013binary,mobilia2003does}. In the context of organizational psychology models, the opposite appears to be true. We will argue that malleable people who have low $A_i$ and thus high natural fluctuations in personality from test to test, not stubborn people who have large $A_i$ and definitive personalities, shape the organization's behavior. This makes intuitive sense if you realize that malleable people are swayed by top-down directives more than stubborn people, and thus can be influenced far more than a stubborn person.

Mathematically, this realization comes from simply staring at the form of Eq. \ref{eq:mu2} and Eq. \ref{eq:Sigma2} and recognizing that $A$ is a measure of innate malleability. If $A$ is large, then naturally the person has little variability in their personality tests when measured in isolation and is stubborn rather than malleable. But if $A$ is small, then there is naturally large variability in their personality tests when measured in isolation, and then such a person is malleable and susceptible to top-down directives.

In the limit that the group or organization is weakly coupled, $\langle A^{-1}\rangle\hat{B}$ is a small correction to $\mathbf{I}_5$, and
\begin{equation}
    \mu\rightarrow -\frac{1}{2}\langle A^{-1}\vec{h}\rangle+\frac{1}{2}\langle A^{-1}\rangle\hat{B}\langle A^{-1}\vec{h}\rangle
\end{equation}
and $\Sigma\rightarrow\Sigma_0$. The dominant term in the mean collective personality comes from people with large $A^{-1}$, meaning people who are malleable. These people also control the fluctuations from $\Sigma$.

Malleable people only control collective personality when the interpersonal coupling is weak. When the organization is strongly coupled, the malleable people do \emph{not} control the group. If we take the limit of $\hat{B}$ large in Eq. \ref{eq:mu2}, we find that $\mu\rightarrow -\frac{1}{2}\hat{B}^{-1}\langle A^{-1}\rangle^{-1}\langle A^{-1}\vec{h}\rangle$, meaning that the effect of malleability from $A$ cancels out. Instead, what matters is the interpersonal coupling itself. Meanwhile, in this limit, $\Sigma\rightarrow -\frac{1}{2N}\hat{B}^{-1}$, meaning that interpersonal coupling controls the fluctuations in collective personality when interpersonal coupling is strong, as expected.

\subsection{Optimal top-down directives}

Suppose we have a group for whom Eqs. \ref{eq:mu2} and \ref{eq:Sigma2} are good approximations to collective personality, and we wish to drive the group towards a particular collective personality $\mu^*,~\Sigma^*$ by means of changing $\hat{B}$ and $\vec{h}_{ext}$ to $\hat{B}^*,~\vec{h}_{ext}^*$, assuming that all external magnetic fields are identical. (In other words, everyone receives the same top-down directive.) We can simply solve for the desired $\vec{h}_{ext}$ and $\hat{B}$:
\begin{eqnarray}
\mu^* &=& -\frac{1}{2}(\mathbf{I}_5+\langle A^{-1}\rangle\hat{B}^*)^{-1}\langle A^{-1}(\vec{h}_{int}+\vec{h}_{ext}^*)\rangle \\
\Sigma^* &=& -\frac{1}{2N} (\hat{B}^*+\langle A^{-1}\rangle^{-1})^{-1}.
\end{eqnarray}
If we rewrite the latter equation, we find that
\begin{equation}
\hat{B}^* = (-2N\Sigma^*)^{-1}-\langle A^{-1}\rangle^{-1}
\end{equation}
and
\begin{eqnarray}
\mu^* &=& -\frac{1}{2}(\mathbf{I}_5+\langle A^{-1}\rangle\hat{B}^*)^{-1}\langle A^{-1}\vec{h}_{int}\rangle \nonumber \\
&& -\frac{1}{2}(\mathbf{I}_5+\langle A^{-1}\rangle\hat{B}^*)^{-1}\langle A^{-1}\rangle\vec{h}_{ext} \\
\rightarrow \vec{h}_{ext} &=&-\langle A^{-1}\rangle^{-1}\langle A^{-1}\vec{h}_{int}\rangle \nonumber \\
&& -2\langle A^{-1}\rangle^{-1}(\mathbf{I}_5+\langle A^{-1}\rangle \hat{B}^*)\mu^*.
\end{eqnarray}
The desired top-down order first cancels out innate personalities and then constructs the desired organizational personality in a way that maps onto everyone's individual personality malleability. Meanwhile, the desired (weak) coupling estimates the deviation between the desired fluctuations in collective personality and the innate fluctuations in collective personality with no coupling and corrects.

\subsection{Collective personality disorders}
\label{sec:personalitydisorders}

From $\mu$ and $\Sigma$, we can understand if a group has a collective personality disorder. Psychopaths are low on agreeableness, neuroticism, and conscientiousness. This means that the second, fourth, and fifth elements of $\mu$ for a collectively psychopathic group will be low. Narcissists are low on agreeableness and high on extraversion, and have fluctuating scores on neuroticism and openness depending on the type of narcissist. This means that the third element of $\mu$ will be high and the fourth element of $\mu$ will be low for a collectively narcissistic group, while the diagonal of $\Sigma$ should reveal relatively large values for the first and fifth dimension.

We wish to see if collective personality disorders are generic. To simplify our search, we assume that the people are homogeneous so that $A$ is the same for all people in the group. When randomly searching through the space of possible $A^{-1}$, $\hat{B}$, $\vec{h}_{int}$, and $\vec{h}_{ext}$ as shown in the Google Colab linked to below, it is not easy to find a combination of these parameters that leads to a valid Maximum Entropy model that indicates a collective personality disorder. However, psychopathic groups can be seen even if all members individually have no personality disorders. Narcissistic groups are harder to find. Group coupling can also work to the group's advantage and reveal a collectively neurotypical group even if individually, the members are psychopathic or narcissistic. I invite readers to modify and randomly generate groups using the code provided in any way they wish, as my choice to use normally distributed random numbers in coupling matrices is completely arbitrary.

\section{Conclusions}

In this manuscript, I have endeavored to tackle a very squishy problem in physics: how to model people in a group, and then more to the point, how to then use this model to ascertain the collective personality of the group. This investigation resulted in a mathematization of ideas about collective intelligence \cite{woolley2010evidence} and collective performance \cite{barrick1991big} proposed in organizational psychology. I proposed that a Maximum Entropy model could be made of the Big Five personality scores of a group of individuals, and that from this Maximum Entropy model, a simple averaging procedure would yield the collective personality of the group. This is an extension of the Curie-Weiss model into a regime that may be more relevant to human thought and behavior. In general, a non-Gaussian Maximum Entropy model could be used, and in those cases, the same idea applies: collective personality emerges as the result of repeated decimation, meaning that the final personality is the final point of a renormalization group flow. The appropriate minimal model should be chosen with expert knowledge, as is true for, say, Maximum Entropy models of neurons \cite{meshulam2025statistical}.

These personality models could be extended to more organisms than just humans. There is no reason a Maximum Entropy model of personalities could not be made of a group of monkeys, for instance \cite{flack2006policing,daniels2017control}. But with humans, it is clear how such a personality model could be made from straightforward measurements, so we have refrained from speculating about the very interesting group dynamics of other social creatures.

Simple qualitative findings emerge just from a straightforward analysis of a Gaussian model of personalities. It turns out that: weak interpersonal coupling is relevant to group personality, a result that has been found in other contexts \cite{schneidman2006weak,ellis1978statistics} and also in the context of groups in a form known popularly as ``the strength of weak ties'' \cite{granovetter1973strength}; malleable people are crucial in determining group personality, against what is found in opinion dynamics models \cite{yildiz2013binary,mobilia2003does}; optimal strategies for directing people in the group so that the group has a certain collective personality can be determined, with directives that can be contrasted with existing literature on qualitatively optimal culture changes \cite{schabracq2007changing}; and collective personality disorders can emerge even if everyone in the group is neurotypical while collective personalities can be neurotypical even if people in the group have personality disorders. More results are likely to emerge simply from understanding different regimes, such as when the Woodbury identity cannot be applied.

I hope that these static results augment the many available opinion dynamics models and group decision-making models \cite{shirzadi2025opinion,kugler2012groups}. I would like to especially note the distinction between the concepts in this manuscript and in those many previous works on opinion formation and group decisions. Nothing in this model is about forming opinions or making decisions, but simply about how one can view the group personality that then might make a wide variety of decisions. The innate character of the group should be relevant to opinion dynamics formation in some interesting and nonlinear way \cite{barrick1998relating}. As a result of the necessary relationship between group character and group decision-making, and of the necessary underlying similarities in mathematical structure because interpersonal correlations in personality usually correspond to correlations in the importance of one person's opinion to the other, there are some similarities in qualitative results about the importance of weak interpersonal coupling.

This work also points towards measurements that could easily be made of group personalities, things that are commonly assessed in practice for team-building exercises in organizations, and straightforward analyses to do thereafter. The Maximum Entropy model suggested here is far easier to fit with data than many of the available dynamic decision-making models, since it can be formed with just an assessment of means and covariances measured with even just five different Big Five personality tests. And unlike spin-glass Ising models, the Lagrange multipliers can be obtained from means and covariances in closed-form \cite{meshulam2025statistical}.

In the future, similar models and analyses could form the basis of a science of collective intelligence \cite{woolley2010evidence}, both intelligence quotient (IQ) and emotional quotient (EQ), and not just collective personalities, although the openness trait in the Big Five personality dimensions correlates with intelligence \cite{barrick1991big}. For instance, IQ and EQ scores could be simply appended to the vectors considered here, revealing when a group of smart people somehow behave stupidly in the collective. Many of the conclusions-- like that weak coupling matters to the collective and that malleable people are more important than stubborn ones-- will hold for such extensions to the model. We leave extended analyses for future research.

\begin{acknowledgments}
We wish to acknowledge the use of Claude in helping with the Woodbury identity and subsequent matrix manipulations. All code is available \href{https://colab.research.google.com/drive/1VE1O6TYjwhHZZQCsQSxrNLk7WsQZLPeY?usp=sharing}{here}. Research was sponsored by the Army Research Office and was accomplished under Grant Number W911NF-25-1-0260. The views and conclusions contained in this document are those of the authors and should not be interpreted as representing the official policies, either expressed or implied, of the Army Research Office or the U.S. Government. The U.S. Government is authorized to reproduce and distribute reprints for Government purposes notwithstanding any copyright notation herein.
\end{acknowledgments}

\appendix
\widetext

\section{Homogeneous populations and the Woodbury identity}
\label{app:Woodbury}

\subsection{Woodbury formula}

In order to find closed-form expressions for $\mu,~\Sigma$ in the homogeneous case, we need to first find a workable expression for $\textbf{J}^{-1}$. The coupling matrix $\textbf{J}$ is a block matrix with $A_i$ on the diagonal and $B$ on off-diagonal. That makes it amenable, when inverting $\textbf{J}$, to using the Woodbury identity.

The Woodbury identity states that for matrices $\mathbf{A} \in \mathbb{R}^{n\times n}$, $\mathbf{U} \in \mathbb{R}^{n\times k}$, $\mathbf{C} \in \mathbb{R}^{k\times k}$, and $\mathbf{V} \in \mathbb{R}^{k\times n}$, if $\mathbf{A}$, $\mathbf{C}$, and $\mathbf{C}^{-1} + \mathbf{V}\mathbf{A}^{-1}\mathbf{U}$ are all invertible, then $\mathbf{A} + \mathbf{U}\mathbf{C}\mathbf{V}$ is invertible and
$$(\mathbf{A} + \mathbf{U}\mathbf{C}\mathbf{V})^{-1} = \mathbf{A}^{-1} - \mathbf{A}^{-1}\mathbf{U}\left(\mathbf{C}^{-1} + \mathbf{V}\mathbf{A}^{-1}\mathbf{U}\right)^{-1}\mathbf{V}\mathbf{A}^{-1}$$
which means that if we identity low-rank corrections to a more-easily diagonalizable matrix in $\textbf{J}$, we can use the Woodbury identity.

Define
$$\tilde{A}_i \equiv A_i - B,~\tilde{\mathbf{A}} \equiv \text{diag}(\tilde{A}_1,\ldots,\tilde{A}_N).$$
If we write
$$\mathbf{J} = \text{diag}(A_1,...,A_N) + (\mathbf{1}\mathbf{1}^\top - \mathbf{I}_N)\otimes B$$
we can rewrite this as
\begin{eqnarray}
\textbf{J} &=& \tilde{\mathbf{A}} + \mathbf{1}\mathbf{1}^\top \otimes B \\
&=& \tilde{\mathbf{A}} + (\mathbf{1}\otimes\mathbf{I}_5)\,B\,(\mathbf{1}^\top\otimes\mathbf{I}_5)
\end{eqnarray}
matching the general form $\mathbf{UCV}$ with
$$\mathbf{A}\to\tilde{\mathbf{A}},\qquad \mathbf{U}\to(\mathbf{1}\otimes\mathbf{I}_5),\qquad \mathbf{C}\to B,\qquad \mathbf{V}\to(\mathbf{1}^\top\otimes\mathbf{I}_5).$$
Then, using the Woodbury formula, we have
\begin{equation}
    \mathbf{J}^{-1} = \tilde{\mathbf{A}}^{-1} - \tilde{\mathbf{A}}^{-1}(\mathbf{1}\otimes\mathbf{I}_5)\Big[B^{-1} + (\mathbf{1}^\top\otimes\mathbf{I}_5)\tilde{\mathbf{A}}^{-1}(\mathbf{1}\otimes\mathbf{I}_5)\Big]^{-1}(\mathbf{1}^\top\otimes\mathbf{I}_5)\tilde{\mathbf{A}}^{-1}.
\end{equation}
The middle term simplifies because $\tilde{\mathbf{A}}^{-1} = \text{diag}(\tilde{A}_1^{-1},\ldots,\tilde{A}_N^{-1})$, so sandwiching it between $\mathbf{1}^\top\otimes\mathbf{I}_5$ and $\mathbf{1}\otimes\mathbf{I}_5$ just sums the blocks:
$$(\mathbf{1}^\top\otimes\mathbf{I}_5)\tilde{\mathbf{A}}^{-1}(\mathbf{1}\otimes\mathbf{I}_5) = \sum_i \tilde{A}_i^{-1}$$
giving the central $5\times5$ object
$$\mathbf{M} \equiv B^{-1} + \sum_i \tilde{A}_i^{-1} = B^{-1} + \sum_i (A_i-B)^{-1}$$
and
\begin{equation}
\mathbf{J}^{-1} = \tilde{\mathbf{A}}^{-1}-\tilde{\mathbf{A}}^{-1} (\mathbf{1}\otimes\mathbf{I}_5) \mathbf{M}^{-1} (\mathbf{1}^{\top}\otimes\mathbf{I}_5)\tilde{\mathbf{A}}^{-1}.
\end{equation}

\subsection{Collective variance}

Because $P=\textbf{1}^{\top}\otimes\textbf{I}_5$, we have from Eq. \ref{eq:meanS}
\begin{eqnarray}
\Sigma &=& P\mathbf{J}^{-1}P^{\top} \\
&=& -\frac{1}{2N^2}\left((\textbf{1}^{\top}\otimes\textbf{I}_5)\tilde{\mathbf{A}}^{-1}(\textbf{1}\otimes\textbf{I}_5)-(\textbf{1}^{\top}\otimes\textbf{I}_5)\tilde{\mathbf{A}}^{-1}(\textbf{1}\otimes\textbf{I}_5) \mathbf{M}^{-1} (\textbf{1}^{\top}\otimes\textbf{I}_5)\tilde{\mathbf{A}}^{-1}(\textbf{1}\otimes\textbf{I}_5)\right) \\
&=& -\frac{1}{2N^2} \left(\sum_i \tilde{A}_i^{-1}-(\sum_i \tilde{A}_i^{-1})\mathbf{M}^{-1}(\sum_i \tilde{A}_i^{-1})\right)
\end{eqnarray}
and by defining
$$\mathbf{Q} \equiv \sum_i \tilde{A}_i^{-1}$$
we have
\begin{equation}
    \Sigma = -\frac{1}{2N^2}\left(\mathbf{Q}-\mathbf{Q}\mathbf{M}^{-1}\mathbf{Q}\right).
\end{equation}
But we have $$\mathbf{Q} = \mathbf{M}-B^{-1}$$ so that
$$
\mathbf{Q}\mathbf{M}^{-1}\mathbf{Q} = \mathbf{Q}\mathbf{M}^{-1}(\mathbf{M}-B^{-1}) = \mathbf{Q}-\mathbf{Q}\mathbf{M}^{-1}B^{-1}\rightarrow \mathbf{Q}-\mathbf{Q}\mathbf{M}^{-1}\mathbf{Q} = \mathbf{Q}\mathbf{M}^{-1}B^{-1}
$$
which implies
\begin{eqnarray}
\Sigma &=& -\frac{1}{2N^2}\mathbf{Q}\mathbf{M}^{-1}B^{-1} \\
&=& -\frac{1}{2N^2} (\mathbf{M}-B^{-1})\mathbf{M}^{-1}B^{-1} \\
&=& -\frac{1}{2N^2} (B^{-1}-B^{-1}\mathbf{M}^{-1}B^{-1})
\end{eqnarray}
which from the Woodbury identity again becomes
\begin{equation}
\Sigma = -\frac{1}{2N^2} (B+\mathbf{Q}^{-1})^{-1} = -\frac{1}{2N^2} (B+\frac{1}{N}\langle (A-B)^{-1}\rangle^{-1})^{-1}.
\end{equation}

\subsection{Collective mean}

To get the collective mean, we use from Eq. \ref{eq:meanS}
\begin{eqnarray}
\mu &=& -\frac{1}{2N}(\mathbf{1}^{\top}\otimes\mathbf{I}_5)\mathbf{J}^{-1}\mathbf{H} \\
&=& -\frac{1}{2N} \left(\sum_i \tilde{A}_i^{-1}\vec{h}_i - (\sum_i \tilde{A}_i^{-1})(B^{-1}+\sum_i\tilde{A}_i^{-1})^{-1}(\sum_i \tilde{A}_i^{-1}\vec{h}_i)\right) \\
&=& -\frac{1}{2}\left(\mathbf{I}_5-N\langle \tilde{A}^{-1}\rangle (B^{-1}+N\langle \tilde{A}^{-1}\rangle)^{-1}\right) \langle \tilde{A}^{-1}\vec{h}\rangle \\
&=& -\frac{1}{2}\left(B^{-1}+N\langle \tilde{A}^{-1}\rangle-N\langle\tilde{A}^{-1}\rangle\right) (B^{-1}+N\langle\tilde{A}^{-1}\rangle)^{-1} \langle\tilde{A}^{-1}\vec{h}\rangle \\
&=& -\frac{1}{2}B^{-1}(B^{-1}+N\langle \tilde{A}^{-1}\rangle)^{-1}\langle\tilde{A}^{-1}\vec{h}\rangle \\
&=& -\frac{1}{2} (\mathbf{I}_5+N\langle\tilde{A}^{-1}\rangle B)^{-1}\langle\tilde{A}^{-1}\vec{h}\rangle.
\end{eqnarray}

\bibliography{apssamp}% Produces the bibliography via BibTeX.

@PREAMBLE{
 "\providecommand{\noopsort}[1]{}" 
 # "\providecommand{\singleletter}[1]{#1}%" 
}

@book{schabracq2007changing,
  title={Changing organizational culture: The change agent's guidebook},
  author={Schabracq, Marc J},
  year={2007},
  publisher={John Wiley \& Sons}
}

@article{granovetter1973strength,
  title={The strength of weak ties},
  author={Granovetter, Mark S},
  journal={American journal of sociology},
  volume={78},
  number={6},
  pages={1360--1380},
  year={1973},
  publisher={University of Chicago Press}
}

@article{dong2024cognitive,
  title={Cognitive biases can move opinion dynamics from consensus to signatures of transient chaos},
  author={Dong, Emily and Marzen, Sarah},
  journal={American Journal of Physics},
  volume={92},
  number={10},
  pages={801--808},
  year={2024},
  publisher={AIP Publishing}
}

@article{woolley2010evidence,
  title={Evidence for a collective intelligence factor in the performance of human groups},
  author={Woolley, Anita Williams and Chabris, Christopher F and Pentland, Alex and Hashmi, Nada and Malone, Thomas W},
  journal={science},
  volume={330},
  number={6004},
  pages={686--688},
  year={2010},
  publisher={American Association for the Advancement of Science}
}

@article{barrick1998relating,
  title={Relating member ability and personality to work-team processes and team effectiveness.},
  author={Barrick, Murray R and Stewart, Greg L and Neubert, Mitchell J and Mount, Michael K},
  journal={Journal of applied psychology},
  volume={83},
  number={3},
  pages={377},
  year={1998},
  publisher={American Psychological Association}
}

@article{bell2007deep,
  title={Deep-level composition variables as predictors of team performance: a meta-analysis.},
  author={Bell, Suzanne T},
  journal={Journal of applied psychology},
  volume={92},
  number={3},
  pages={595},
  year={2007},
  publisher={American Psychological Association}
}

@article{barrick1991big,
  title={The big five personality dimensions and job performance: a meta-analysis},
  author={Barrick, Murray R and Mount, Michael K},
  journal={Personnel psychology},
  volume={44},
  number={1},
  pages={1--26},
  year={1991},
  publisher={Wiley Online Library}
}

@article{slobodskaya2021personality,
  title={Personality development from early childhood through adolescence},
  author={Slobodskaya, Helena R},
  journal={Personality and Individual Differences},
  volume={172},
  pages={110596},
  year={2021},
  publisher={Elsevier}
}

@article{castellano2009statistical,
  title={Statistical physics of social dynamics},
  author={Castellano, Claudio and Fortunato, Santo and Loreto, Vittorio},
  journal={Reviews of modern physics},
  volume={81},
  number={2},
  pages={591--646},
  year={2009},
  publisher={APS}
}

@article{tkavcik2013simplest,
  title={The simplest maximum entropy model for collective behavior in a neural network},
  author={Tka{\v{c}}ik, Ga{\v{s}}per and Marre, Olivier and Mora, Thierry and Amodei, Dario and Berry II, Michael J and Bialek, William},
  journal={Journal of Statistical Mechanics: Theory and Experiment},
  volume={2013},
  number={03},
  pages={P03011},
  year={2013},
  publisher={IOP Publishing and SISSA}
}

@article{daniels2017control,
  title={Control of finite critical behaviour in a small-scale social system},
  author={Daniels, Bryan C and Krakauer, David C and Flack, Jessica C},
  journal={Nature communications},
  volume={8},
  number={1},
  pages={14301},
  year={2017},
  publisher={Nature Publishing Group UK London}
}

@article{flack2006policing,
  title={Policing stabilizes construction of social niches in primates},
  author={Flack, Jessica C and Girvan, Michelle and De Waal, Frans BM and Krakauer, David C},
  journal={Nature},
  volume={439},
  number={7075},
  pages={426--429},
  year={2006},
  publisher={Nature Publishing Group UK London}
}

@article{ellis1978statistics,
  title={The statistics of Curie-Weiss models},
  author={Ellis, Richard S and Newman, Charles M},
  journal={Journal of Statistical Physics},
  volume={19},
  number={2},
  pages={149--161},
  year={1978},
  publisher={Springer}
}

@article{kugler2012groups,
  title={Are groups more rational than individuals? A review of interactive decision making in groups},
  author={Kugler, Tamar and Kausel, Edgar E and Kocher, Martin G},
  journal={Wiley interdisciplinary reviews: Cognitive science},
  volume={3},
  number={4},
  pages={471--482},
  year={2012},
  publisher={Wiley Online Library}
}

@article{shirzadi2025opinion,
  title={Opinion dynamics: A comprehensive overview},
  author={Shirzadi, Mohammad and Cruciani, Emilio and Zehmakan, Ahad N},
  journal={arXiv preprint arXiv:2511.00401},
  year={2025}
}

@article{mobilia2003does,
  title={Does a single zealot affect an infinite group of voters?},
  author={Mobilia, Mauro},
  journal={Physical review letters},
  volume={91},
  number={2},
  pages={028701},
  year={2003},
  publisher={APS}
}

@article{yildiz2013binary,
  title={Binary opinion dynamics with stubborn agents},
  author={Yildiz, Ercan and Ozdaglar, Asuman and Acemoglu, Daron and Saberi, Amin and Scaglione, Anna},
  journal={ACM Transactions on Economics and Computation (TEAC)},
  volume={1},
  number={4},
  pages={1--30},
  year={2013},
  publisher={ACM New York, NY, USA}
}

@article{van2025coding,
  title={Coding schemes in neural networks learning classification tasks},
  author={van Meegen, Alexander and Sompolinsky, Haim},
  journal={Nature Communications},
  volume={16},
  number={1},
  pages={3354},
  year={2025},
  publisher={Nature Publishing Group UK London}
}

@article{schneidman2006weak,
  title={Weak pairwise correlations imply strongly correlated network states in a neural population},
  author={Schneidman, Elad and Berry, Michael J and Segev, Ronen and Bialek, William},
  journal={Nature},
  volume={440},
  number={7087},
  pages={1007--1012},
  year={2006},
  publisher={Nature Publishing Group UK London}
}

@article{lee2015statistical,
  title={Statistical mechanics of the US Supreme Court},
  author={Lee, Edward D and Broedersz, Chase P and Bialek, William},
  journal={Journal of Statistical Physics},
  volume={160},
  number={2},
  pages={275--301},
  year={2015},
  publisher={Springer}
}

@article{bialek2012statistical,
  title={Statistical mechanics for natural flocks of birds},
  author={Bialek, William and Cavagna, Andrea and Giardina, Irene and Mora, Thierry and Silvestri, Edmondo and Viale, Massimiliano and Walczak, Aleksandra M},
  journal={Proceedings of the National Academy of Sciences},
  volume={109},
  number={13},
  pages={4786--4791},
  year={2012},
  publisher={National Academy of Sciences}
}

@article{bialek2014social,
  title={Social interactions dominate speed control in poising natural flocks near criticality},
  author={Bialek, William and Cavagna, Andrea and Giardina, Irene and Mora, Thierry and Pohl, Oliver and Silvestri, Edmondo and Viale, Massimiliano and Walczak, Aleksandra M},
  journal={Proceedings of the National Academy of Sciences},
  volume={111},
  number={20},
  pages={7212--7217},
  year={2014},
  publisher={National Academy of Sciences}
}

@article{meshulam2025statistical,
  title={Statistical mechanics for networks of real neurons},
  author={Meshulam, Leenoy and Bialek, William},
  journal={Reviews of Modern Physics},
  volume={97},
  number={4},
  pages={045002},
  year={2025},
  publisher={APS}
}

@article{samuel2008meta,
  title={A meta-analytic review of the relationships between the five-factor model and DSM-IV-TR personality disorders: A facet level analysis},
  author={Samuel, Douglas B and Widiger, Thomas A},
  journal={Clinical psychology review},
  volume={28},
  number={8},
  pages={1326--1342},
  year={2008},
  publisher={Elsevier}
}

\end{document}